\documentclass[twocolumn,aps,showpacs]{revtex4}
\usepackage{graphicx}
\begin{document}
\title{Heteroepitaxial growth of high-K gate oxides on silicon:
insights from first-principles calculations on Zr on Si(001)}
\author{Clemens J. F\"orst,$^{1,2}$ Peter E. Bl\"ochl,$^{1}$ and Karlheinz Schwarz$^{2}$}

\affiliation{$^1$ Clausthal University of Technology, Institute for
Theoretical Physics, Leibnizstr.10, D-38678 Clausthal-Zellerfeld,
Germany}
\affiliation{$^2$ Vienna University of Technology, Institute for
Materials Chemistry, Getreidemarkt 9/165-TC, A-1060 Vienna, Austria}
\begin{abstract}
Metal deposition of Zr an a Si(001) surface has been studied by
state-of-the-art electronic structure calculations.  The energy per Zr adatom
as a function of the coverage shows, that Zr forms  silicide islands even
at low coverages.  Adsorbed Zr is thermodynamically unstable against the
formation of bulk silicide ZrSi$_2$.  The observation that the islands consist
of structural elements of the bulk silicide is an indication that silicide
grains will form spontaneously.
\end{abstract}
\date{\today}
\maketitle
%
\section{Introduction}

The scaling of the CMOS transistor has been the driving force
behind the tremendous increase in microprocessor performance
observed during the last decades.  While the problems of the past
were dominated by manufactural aspects, one now faces the first
fundamental physical limitations as structures in
logical devices approach atomic dimensions.

As a rule of thumb, the thickness of the gate-oxide has to be
directly proportional to the channel length in a MOSFET (metal
oxide silicon field effect transistor) device.  In the course of
the ongoing miniaturization, also the thickness of this insulating
layer is being continually reduced.  By the year 2007, gate-oxides
in a transistor will approach a thickness of
1.5\,nm\cite{roadmap}, which corresponds to about ten typical
atomic distances. The quantum mechanical tunneling currents
through such a thin oxide are intolerable, and cause increased
power consumption and deteriorated switching characteristics of
the transistor.  Replacing SiO$_2$-based oxides, nowadays employed
as gate dielectric, is one of the ``key challenges'' to the
semiconductor industry\cite{roadmap}, which has to find a solution
within the next 4 to 5 years.

Employing high-k ($\equiv$ large dielectric constant) oxides would
allow for a greater physical thickness while preserving the
electrical properties.  Possible candidates have to meet an
extensive list of requirements\cite{wil01} such as sufficiently
large band offsets, thermodynamical stability or interface
quality.

The Si--SiO$_2$ system meets all these requirements in an unparallelized
way\cite{buc00}.  First attempts to grow alternative oxides on Si(001) did not
yield satisfactory results for a variety of reasons -- often originating in the
interface region.

In the course of changing to a new gate material, one also considers to now
grow crystalline oxide layers.  It has been a paradigm that the gate oxide
must be amorphous in order to avoid dislocations and grain boundaries, which
provide natural pathways for leakage currents and atom diffusion. However, in
the very small devices used in the near future, the probability for such a
defect may be sufficiently small. On the other hand,
epitaxial oxides bear the promise of very low defect concentrations at the
interface to silicon, which is crucial for efficient device operation.

The first account of heteroepitaxial growth of a transition metal oxide has
been presented by a seminal work of McKee\cite{mck01} who succeeded in growing
SrTiO$_3$ on Si(001).  For the divalent elements detailed studies have been
performed\cite{yao99,oji01,her00,bak96,fan90,wan99}.  As similar understanding
does not yet exist for the technological relevant early transition metals such
as Zirconium.  Theoretical studies on Zr are limited bulk interface
calculations of silicates\cite{kaw01}.

The deposition and formation of an interface between one of the
major contenders for high-k oxides, namely Zirconia (ZrO$_2$), and
silicon has not yet been investigated with ab-initio simulations.
As a first step we performed state-of-the-art electronic structure
calculations of the deposition of Zr on Si(001) up to a coverage
of two monolayers (ML).

\section{Computational Aspects}

We performed first-principles calculations within the framework of density
functional theory\cite{dft1,dft2} using the gradient corrected density
functional of Perdew, Burke and Enzerhof\cite{pbe}. The electronic structure
problem has been solved with the projector augmented waves method\cite{blo94},
which uses augmented plane waves to describe the full wave-functions and
densities without shape approximation. The PAW method as implemented in the
CP-PAW code employs the Car-Parrinello approach\cite{cpmd} to minimize the
total energy functional.

The core electrons are described within the frozen core
approximation with the semi-core (4$s$ and 4$p$) shells of the Zr
atoms treated as valence electrons. Plane wave cutoffs of 30 and
60\,Ry for the wave functions and the density have been used.

For metallic system we minimized the Mermin functional with
respect to the occupation numbers which yields the Fermi-Dirac
distribution for the electrons.  An electronic temperature of
1000\,K was used.  The zero-Kelvin result has been extrapolated
using the method suggested by~\cite{sch97}.

The reconstructed Si(001) surface is the template for our growth studies.  It
is modeled by a slab consisting of 5 silicon layers, where the bottom layer is
saturated with two hydrogen atoms per silicon.  The position of the hydrogen
atoms and the lowest silicon layer has been frozen.  Supercells with 16 atoms
per layer have been used for the calculations of Zr in the dilute limit, 8
atoms per layer for all higher coverages.  In the first case, the distance
between two periodic images of an adatom is 15.36\,\AA .  For all surface
calculations the k-mesh density corresponds to 64 lateral k-points per
$p(1\times1)$ surface unit cell.

\section{Dilute Limit}

\begin{figure}[ht]
\includegraphics[angle=90,draft=false,width=0.45\textwidth]{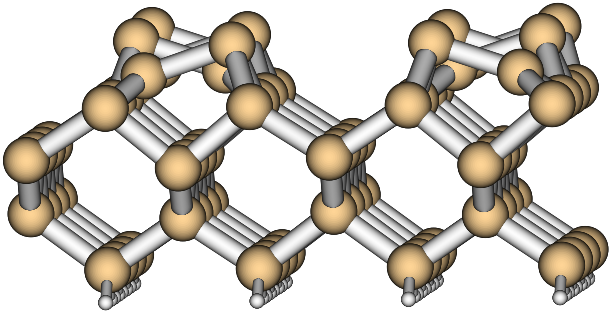}
\caption{Si(001) surface with $c(4\times2)$ reconstruction. The figure
represents one supercell of the slab as used as a template in our calculations.}
\label{fig1}
\end{figure}

Fig.~\ref{fig1}. shows the structure of the reconstructed Si(001) surface.  Excellent
accounts covering the full complexity of the surface reconstruction can be
found in literature\cite{cha79,sch01}.  The resulting structure can, however,
be explained by the following considerations: While every bulk silicon is
4-fold coordinated, the surface atoms lack their upper bonding partners which
leaves 2 dangling bonds per atom, each occupied by one electron.  In a first
step, two neighboring silicon atoms form a dimer bond -- initially parallel to
the surface -- which leaves one dangling bond per atom. Now both electrons are
transferred into one dangling bond to reduce the number of unpaired electrons
to zero.  This results in the buckling as the silicon with the filled dangling
bond prefers a tetrahedral bonding arrangement, while the other prefers a
planar sp$^2$ configuration.  From the electrostatic point of view, the
alternating buckling behavior is favorable.  Our calculations readily reproduce
previously reported results such as the $c(4\times2)$ reconstruction and the
difference in $z$-coordinate of two silicon atoms within a dimer\cite{nor93}.

The energy surface of a Zr adatom on Si(001), has been obtained by freezing the lateral
position of the adsorbed atom relative to the slab backplane, while all other
degrees of freedom were fully relaxed. We used a grid of twelve Zr
positions in the irreducible zone of the reconstructed $p(2\times1)$ silicon surface.

\begin{figure}[ht]
\includegraphics[draft=false,width=0.45\textwidth]{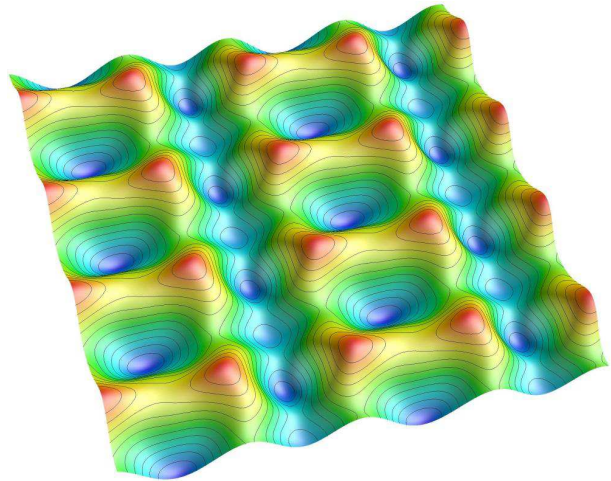}
\caption{The total energy surface of an isolated Zr adatom as a function of the
lateral position on the Si(001) surface.  The dimer silicon atoms are located
at the bar-shaped maxima on this surface.  The valleys in the energy map
correspond to the valleys between the dimer rows.}
\label{fig2}
\end{figure}

The resulting total energy surface for an isolated Zr on top of
Si(001) is shown in Fig.~\ref{fig2}. Two nearly degenerate positions can be
identified.  One is located in the valley right in the middle of
two dimers of neighboring rows and the other on top of a dimer row
between to adjacent dimers.  A third, local minimum, is located in
the valley and has an energy 0.30\,eV higher relative to the most
favorable positions. The diffusion is quasi one-dimensional with
barriers of 0.70\,eV for diffusion parallel to the dimer rows and
1.63\,eV from the valley to the row.

\section{Coverage of 0.25 and 0.5 Monolayers}

\begin{figure}[ht]
\includegraphics[width=0.45\textwidth,draft=false,clip=true]{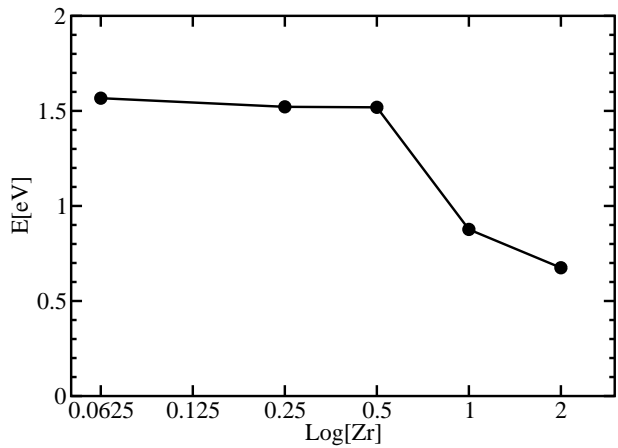}
\caption{ The energy per Zr adatom as a function of the coverage.
Energies are given relative to bulk ZrSi$_2$ and bulk Si.}
\label{fig3}
\end{figure}

In order to investigate the formation of a continuous film, we increased the
coverage. The structures with coverages of 0.25 ML and 0.5 ML have similar
energies per Zr atom as the dilute limit (see Fig.~\ref{fig3}). A wealth of complex structures has
been found. Here, however, we only summarize the main trends of the chemical
binding.

To first approximation Zr prefers a formal 4+ charge state -- the projected
density of states of the Zr $d$-states is located well above the Fermi
level. It is well known that for transition metal cations the $s$-and
$p$-electrons are located above the \mbox{$d$-states}, and have only a minor
effect on the occupied states.  The silicon dimers at the surface accept up to
two electrons in their dangling bonds. A clear structural indication that the
dangling bonds are filled is the disappearance of the dimer buckling.

Further electrons that are supplied at increased coverage of Zr, occupy the
antibonding states of the dimer bond.  As a consequence the silicon dimers
break up. This happens at a coverage of half a ML and above. It should be
noted, however, that metallic states accept some of the electrons so that the
number of broken dimers does not directly correlate with the number of Zr
adatoms in the ratio one to one.  As a result of the interplay between breaking
dimers and metallic states we find fairly complex reconstructions of the
surface structure for intermediate coverages.

\section{Silicide Formation}

\begin{figure}[ht]
\includegraphics[angle=90,width=0.45\textwidth,draft=false,clip=true]{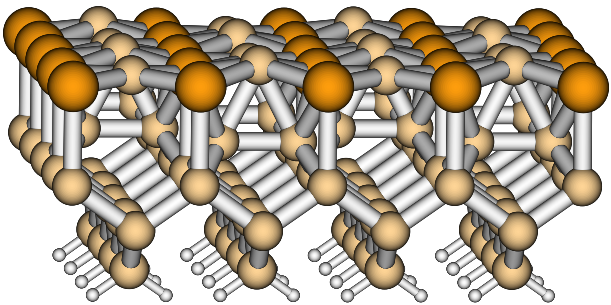}
\caption{ The structure of a Zr monolayer.  Below the silicide layer, subsurface
dimers have formed.}
\label{fig4}
\end{figure}

Upon increase of the coverage to a full ML, the energy
drops by 0.64 eV per adatom (see Fig.~\ref{fig3}).  The stable structure with ML
coverage is shown in Fig.~\ref{fig4}.
All dimer bonds of the surface layer are broken, and the Zr atoms
occupy the centers of the square array of surface silicon atoms in
the resulting $p(1\times1)$ reconstruction. Such a layer is one
structural element of bulk ZrSi$_2$.

The energy gain is, however, not due to the surface geometry but can be
attributed to a dimer reconstruction of the silicon subsurface. A metastable
state without this reconstruction is even higher in energy than adsorbed Zr at
lower coverages.

During the reconstruction, the silicon atoms in the layer underneath the ZrSi
surface layer form dimers analogous to the bare silicon surface. In contrast to
the dimer row reconstruction of the silicon surface, these subsurface dimers
are not buckled and are arranged in a checkerboard instead of a row pattern.

\begin{figure}[ht]
\includegraphics[angle=90,width=0.45\textwidth,draft=false,clip=true]{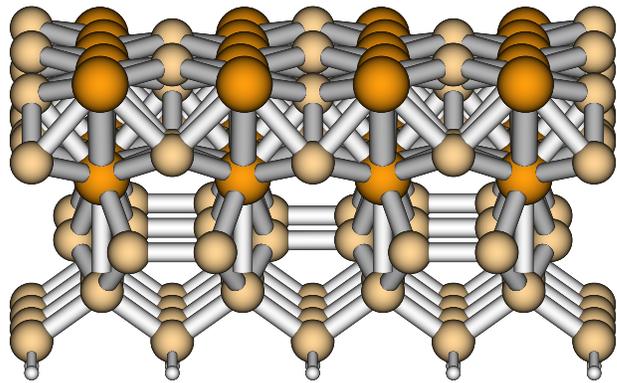}
\caption{ Structure obtained at a structure of two monolayers of Zr.  The ZrSi
 double layer constitutes a structural element of bulk ZrSi$_2$ (compare to
 Fig.~6).  Note the dimer reconstruction of the silicon layer underneath the
 ZrSi double layer.  }
\label{fig5}
\end{figure}

\begin{figure}[ht]
\includegraphics[angle=90,width=0.45\textwidth,draft=false,clip=true]{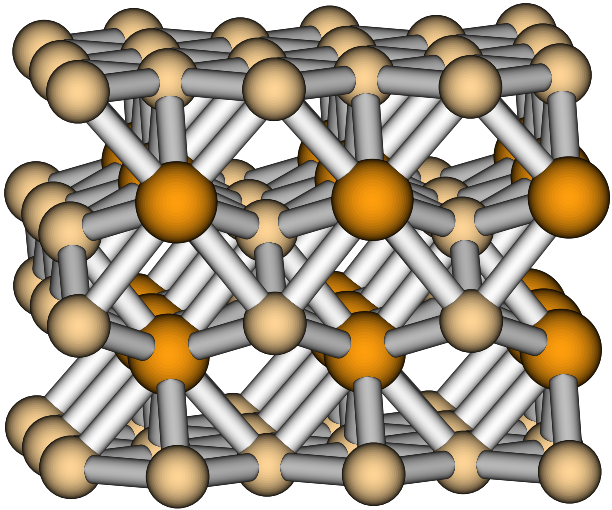}
\caption{ Structure of ZrSi$_2$ bulk.  It consists of pure silicon
layers separated by a ZrSi double layers.  This double layer
exhibits Si zig-zag chains separated by Zr atoms.  }
\label{fig6}
\end{figure}

At a coverage of two MLs a second ZrSi layer is formed.  This configuration is
shown in Fig.~\ref{fig5}. The second layer is nearly identical to the first, but is
shifted laterally. The double layer is again a structural element of bulk
ZrSi$_2$, the structure of which is shown in Fig.~\ref{fig6}. The ZrSi double layer
consists of one-dimensional Si zig-zag chains separated by Zr atoms. The Zr
atoms lie approximately in the plane of the upper and lower atoms of the Si
chain.  As for ML coverage the first silicon layer underneath the ZrSi double
layer exhibits a dimer reconstruction.

Despite an additional energy gain from the ML coverage to a coverage of two MLs
of 0.25\,eV per adatom, even this structure is still 0.67 eV higher in energy
than bulk ZrSi$_2$.  While the atomic process has not yet been resolved in
every detail, our findings give strong indications for the nucleation of the
silicide.

\section{Discussion and Conclusion}

We presented the results of state of the art ab-inito electronic structure
calculations aiming at understanding the deposition of Zr atoms on a Si(001)
surface as it is the case in an MBE reaction chamber.

Our results are summarized in Fig.~\ref{fig3}, which shows the energy per Zr adatom as a
function of the coverage.  For coverages below one monolayer the the energy is
nearly independent of the coverage. At a coverage of 1 ML we observe a sharp
drop in energy by 0.64\,eV, followed by further drops in energy for higher
coverages. All structures are less stable than bulk silicide.

Our findings suggest that islands with a local coverage of 1 ML or
higher are formed even at low coverages. The islands contain
structural elements of bulk ZrSi$_2$, which is more stable than
any surface structure. Therefore a likely scenario is the
formation of bulk silicide grains, that disrupt the surface
morphology, and are detrimental for epitaxial growth. Modification
of growth conditions, such as exposing the surface to a oxygen
containing ambient, may bypass silicide formation during the first
growth steps.

\section*{Acknowledgments}

This work has been funded by the European Commission in the
project \mbox{"INVEST"} (Integration of Very High-k Dielectrics
with CMOS Technology) IST-2000-28495 and by the AURORA project
(\mbox{SBF F011}) of the Austrian Science Fond. This work has
benefited from the collaborations within the ESF Programme on
'Electronic Structure Calculations for Elucidating the Complex
Atomistic Behavior of Solids and Surfaces'.

\bibstyle{plain}
\bibliographystyle{plain}

\end{document}